# Modeling Escape from a One-Dimensional Potential Well at Zero or Very Low Temperatures

**Chungho Cheng**

*Department of Mechanical and Aerospace Engineering, University of California, Davis, California 95616, USA*

**Gaetano Salina**

*Istituto Nazionale di Fisica Nucleare, Sezione Roma "Tor Vergata", I-00133 Roma, Italy*

**Niels Grønbech-Jensen**

*Department of Mathematics and Department of Mechanical and Aerospace Engineering, University of California, Davis, California 95616 , USA*

**James A. Blackburn**

*Physics & Computer Science, Wilfrid Laurier University, Waterloo, ON N2L 3C5, Canada*

**Massimiliano Lucci and Matteo Cirillo (\*)**

*Dipartimento di Fisica and MINAS-Lab Università di Roma "Tor Vergata", I-00133 Roma, Italy*

## Abstract

The process of activation from a one-dimensional potential is systematically investigated in zero and nonzero temperature conditions. The features of the potential are traced through statistical escape from its wells whose depths are tuned in time by a forcing term. The process is carried out for the damped pendulum system imposing specific initial conditions on the potential variable. While the escape properties can be derived from the standard Kramers theory for relatively high values of the dissipation, for very low dissipation these deviate from this theory by being dependent on the details of the initial conditions and the time dependence of the forcing term. The observed deviations have regular dependencies on initial conditions, temperature, and loss parameter itself. It is shown that failures of the thermal activation model are originated at low temperatures, and very low dissipation, by the initial conditions and intrinsic, namely *T=0*, characteristic oscillations of the potential-generated dynamical equation.

(\*) Corresponding author : cirillo@roma2.infn.it

# 1) INTRODUCTION

The interest for fluctuations in dynamical systems and the analysis of their dynamical response when subject to external forcing terms and/or thermal noise is a subject that has attracted generations of scientists. Several relevant reviews are available on this topic [1, 2, 3] and body of work and insight are still expanding in part due to increased power of numerical techniques. A key model for describing thermal activation is due to Kramers [4,5]; a relevant feature of this model is that the escape rate $\Gamma$ from a potential well with amplitude $\Delta U$, at temperature $T$, is governed by the equation [4]

$$\Gamma = f \exp(-\frac{\Delta U}{k_B T}) \qquad (1)$$

where $f$ is an attempt frequency. A system to which Kramers model (and eq. 1) has been extensively applied is the compound pendulum, which is described by the following normalized equation for the angle $\varphi$

$$\ddot{\varphi} + \alpha\dot{\varphi} = -\frac{dU}{d\varphi} \qquad (2)$$

with $U(\varphi) = (1 - cos\varphi) - \eta\varphi$ being the potential energy. The coefficient $\alpha$ and the parameter $\eta$ account respectively for dissipation and forcing terms. Two recent reviews [6,7] demonstrate the relevance of the system (2) in the history of physics and of show its intriguing counterparts in general physics and nonlinear dynamics. In condensed matter the driven-damped pendulum equation is analogous to that of the Josephson junction [8], which has been the focus of much attention in theoretical, experimental, and applied physics level for decades [9,10]. Specific applications of Kramers analysis to Josephson potentials were reported in two key publications by Kurkijarvi [11] and Büttiker et al. [12].

Due to the applicability of Kramers model to a broad class of systems, we have conducted a systematic statistical analysis of its validity for describing the behavior of the pendulum. We first study, in the next section, the potential in the absence of temperature fluctuations, i. e. studying eq.

2 in absence of noise. In this *T=0* case we focus on the dependence of the escape from the potential for different values of dissipation values while increasing the force term linearly on time. We study the effect of specific initial conditions on the statistical distributions of the escape processes analyzing the response induced by the initial data. We also present a model which explains the underlying physics of the obtained numerical results. Then, in sect. 3, we include thermal effects through a noise term in eq. 2. We show that the conclusions made for the zero-temperature case about the initial conditions can play a crucial role for the escape statistics in regime of low dissipation and nonzero temperature. In sect. 4 we discuss the results in terms of experimental phenomenology while in sect. 5 we summarize the paper.

## 2) ZERO TEMPERATURE

In Fig. 1a the curves show the first well of the potential $U$ that we are investigating traced for increasing values of the parameter $\eta$ ($\eta = 0.6$, the lowest and $\eta = 1$ the highest). The inset shows the value of the height of the potential barrier $\Delta U$, the energy spacing between the maximum and minimum value in the well, spanned over the entire interval $[0;1]$ of the forcing term $\eta$ ; according to the dependence $\Delta U = 2(\sqrt{1-\eta^2} - \eta \cos^{-1}\eta)$, the depth of the potential (the difference in energy between successive minimum and the maximum) decreases and goes to zero for $\eta =1$ as shown in the inset. In our escape experiments and numerical procedures the forcing term is increased according to $\eta(t) = \dot\eta t$ where the derivative $\dot\eta = \frac{d\eta}{dt}$ is kept constant and the curves in Fig. 1a, in practice, correspond to successive time shots for $\dot\eta = $ 1.95 x 10$^{-8}$. In the figure we also indicate the stable equilibrium points at the bottom of the potential (full diamonds) and the unstable (empty circles) equilibrium points at the top of the well: these were calculated, for each curve evaluating numerically the minima and the maxima.

In Fig. 2 we trace both stable and unstable equilibrium points as a function of $\eta$, respectively by the continuous and the dashed lines: the crossing of the two curves is the point where an escape event is recorded: we see that this event occurs, always when $\eta =1$, when the well becomes flat and the two curves cross, meaning that the stable and unstable equilibrium point have the same $(\eta,\varphi)$ coordinates. The escape event is numerically identified by the fact that, for a further increase of the forcing term above $\eta=1$, the system is driven in a dynamical condition of continuous phase increase (rotation states in terms of the compound pendulum). For $T=0$ (zero temperature) and initial conditions $\varphi_0 = \varphi(t=0) = 0$ and $\dot\varphi_0 = \frac{d\varphi}{dt}(t=0) = 0$, the escape from the potential well always occurs for $\eta=1$; moreover, for these initial conditions, this situation remains the same, for any value of $\alpha$ and $\eta$. The two plots in Fig. 2a,b represent indeed the results of the numerical integration for these initial conditions and different values of $\alpha$ (indicated in the plot) : we can see that, even decreasing the loss parameter of six orders of magnitude, the response is always the same and the escape always occurs above $\eta =1$.

Before proceeding, however, it is necessary to specify what is our criterion to judge that an escape process out of the potential well has safely occurred. When $\eta < 1$ the escape occurs at the angle $\varphi = \pi - \arcsin \eta$ while for $\eta \approx 1$ the escape is recorded when $\varphi = \frac{\pi}{2}$. In both cases the escape from the well is double checked by the continuous increase of the phase generating a non–zero average value of the time derivative of the phase (we usually stop the integration when $<\dot\varphi> = 1$). The results herein presented have been obtained integrating the system (2) by specific versions of the velocity-explicit Størmer-Verlet algorithm [13] setting the integration time $\Delta t = 0.02$ ; two independently developed versions of the algorithm, running on different computers and operating systems, always returned the same results. Halving tests for the integration time step were performed routinely in order to check the independence of the results upon it.

Keeping the same value of $\dot\eta = 1.95 x 10^{-8}$ but setting nonzero values for the initial angle $\varphi_0$ the dependence on the loss parameter of the escape process becomes significant and the values of $\eta$ for which escape from the potential well are recorded can be substantially different. Fig. 3 and Fig. 4 show two escape processes obtained both for initial condition $\varphi_0 \neq 0$ (respectively $\varphi_0 =0.3\pi$ and $\varphi_0 =0.6$ $\pi$ for Fig. 3 and Fig. 4) and three decreasing values of the loss parameter $\alpha$. The value of time derivative of the initial phase $\dot\varphi_0$ was set to zero. The physical meaning of these initial conditions for the system (2) corresponds to start an integration in an oscillating regime with zero average phase derivative in the point where the phase has its maximum value and the derivative of the phase is zero, namely the inversion point of the oscillations. We see now that in Fig. 3a,b and 4a,b the "phase trajectory" still escapes from the potential well for $\eta$ values very close to the critical one (the unity), however, for $\alpha$ in the $10^{-7}$ range the phase oscillations due to the initial angle do not damp out and the dark area in the figures is just generated by the oscillations whose period is too small to be seen in the figures (we just show a zoom with few periods before escaping in the inset of Fig. 4b). In this case the escape occurs, as shown in the figure, when the phase oscillation trajectory crosses the unstable point trace for $\eta <1$ . Considered the effect shown in Fig. 3 and Fig. 4, we decided to investigate systematically the effect of dissipation and initial conditions on the escape current, namely the value of the bias current for which the system escapes from the well.

The result of our simulations for the dependence of the escape current upon loss parameter and initial angle are epitomized in Fig. 5. Here we show in a three-dimensional plot the data obtained for the escapes out of the potential well of the system (2) for $T=0$ setting $\dot{\eta} = 1.95 \times 10^{-8}$ . On the vertical axis we report the observed value of the "escape" of $\eta_E$, namely the value of $\eta$ generating an escape event from the periodic oscillations in the potential well to an out of the well running state in which the angle increases continuously. On the horizontal axes we have the initial angle $\varphi_0$ which is indicated in the figures as $\varphi_0$ and the loss parameter $\alpha$ . We can see that when the initial angle tends to zero the escape always occurs for $\eta_E=1$ , for any value of the loss parameter. Instead, when the loss decreases and the initial angle increases we see that the escape current decreases substantially until it reaches zero actually when $\varphi_0 = \pm\pi$ and values of the loss parameter less than $10^{-7}$. Changing the value of the parameter $\dot{\eta}$ of one order of magnitude no substantial differences are observed in the escape plots of Fig. 5: we always see that relevant changes occur below $log\alpha \cong -7$ .

In Fig. 6 we show two section of plots like that of Fig. 5 where we can see more clearly the effect of the loss parameter condition (a) and initial conditions (b): these are used as parameters for the curves in the figures where, on the vertical axis we always report the value of the escape current. Both these "sections" were obtained for $\dot{\eta} =1.95 \times 10^{-8}$. In (a) we can clearly see that the escape is very flat down to $10^{-7}$ and then it becomes a monotonically decreasing and symmetric function of the angle . Decreasing the loss, the width of the bell-shaped function shrinks but saturates for $log\alpha =-9$. In (b) we show the dependence of the escape current upon the parameter $\alpha$ for different initial conditions : here we plot the escape current versus the $log\ \alpha$ and we parametrize with respect to the initial angle and the result of this operation is just to set the value of a specific escape current on the left side of the plot while on the right side, for $\alpha > 10^{-7}$, the escape current is always equal to the unity.

We then investigated the response of the system by changing the parameter $\dot{\eta}$ for a given initial angle, $\varphi_0 = 0.2\,\pi$ and different values of the loss parameter. The results are shown in Fig. 7: here in (a) we see that variations in $\dot{\eta}$ even of one order of magnitude, generate slight differences in the escape currents around the loss values of the order of $10^{-7}$; the interesting thing now is to consider what happens if we scale the horizontal axis of the data in (a) plotting the escape current as a function of the common logarithm of ratio $\kappa = \frac{\alpha}{\dot{\eta}}$. Now we find (see Fig. 7b) that all the data collapse in a single curve. The data show that $\kappa$, the ratio between loss parameter ($\alpha$) and rate of increase of the forcing term ($\dot{\eta}$), plays an important role providing a sort of normalization of the observed potential escape features and a discontinuity in the response of the system occurs when $\kappa = 1$. The result of Figs. 2-7 summarize our investigations for $T=0$ : given a specific initial condition on the angle coordinate of the potential, the value of $\eta$ for which we observe escape out of the potential well depends in a relevant way from the values of the parameters $\alpha$ and $\dot{\eta}$.

We conclude that the system (2) undergoes an abrupt "transition" when $\kappa$, the ratio between dissipation and rate of increase of applied force, is equal to the unity. This ratio has also been shown to play a relevant role in the context of the appearance of multipeaked escape statistical distributions when sweep rates are particularly high [14,15]. Since we limit the investigation herein to statistical distributions obtained with relatively low sweep rates, multipeaked distributions like those reported in ref. 14 are not observed. As we shall see now the analysis for $T=0$ of the escape process is a relevant physical background for understanding thermal excitations in the well. This is not surprising indeed because it is known that the characteristics of a nonlinear system for $T=0$ leads to identifying relevant phenomena [7] and it has been shown how complex can be the $T=0$ spectrum of "simple" nonlinear systems [16] and how anharmonic analysis of the potentials can provide account for experimental results [17].

In terms of Josephson effect physics the role of the oscillations generated by an initial angle on the escape value $\eta_E$ (the current for which the junctions switch from zero to non-zero voltage state) can be understood, for low dissipation values, on the basis of a model for the energy stored in the Josephson inductance in the zero voltage state. Given the value of the maximum Josephson pairs current $I_c$ the Josephson inductance is defined by [9]

$$L_J = \frac{1}{\sqrt{1-(\frac{I}{I_c})^2}} \frac{\Phi_0}{2\pi I_c} \qquad (3)$$

Where $I$ is the bias current fed through the junction and $\Phi_0 = 2.07 \times 10^{-15}$ $Wb$ is the flux quantum. In the limit $I=0$ the above equation returns the zero-bias Josephson inductance $L_{J0} = \Phi_0/2\pi I_c$.

An initial angle $\varphi_0$ gives rise, according to the Josephson dc equation, to an initial current in the form $I_0 = I_c \sin \varphi_0$. We assume that, for $t > 0$, an oscillating current with amplitude $I_0$ and oscillating time dependence in the form $I_0 \sin \omega_j t$ will be activated. Here we can consider $\omega_j$ as the bias-dependent Josephson angular plasma oscillations frequency [9]. In this harmonic approximation it is straightforward to calculate the average energy in one period of these oscillations in the Josephson inductor as $W_L = \frac{1}{4} L_j I_0^2$.

We presume that, when damping is very low this initial energy will persist in the system even when the value of the forcing term is slowly increased and indeed these oscillations are those that we can see in the inserts of Figs. 3 and 4. Thus, the just calculated $W_L$ will contribute to lower the barrier height $\Delta U$ of the washboard potential in one period in which an escape attempt occurs. We recall now that in the expression for $\Delta U$ given before, in Josephson terms, energies are normalized to the Josephson zero bias energy, namely $E_J = \Phi_0 I_c/2\pi$. Normalizing $W_L$ to this quantity and subtracting it from $\Delta U$ we get an effective height of the potential $\Delta U_E$ in the form :

$$\Delta U_E = 2\left(\sqrt{1-\eta^2} - \eta \cos^{-1}\eta\right) - \frac{\eta_0^2}{4\sqrt{1-\eta_0^2}} \qquad (4)$$

where $\eta_0 = I_0/I_c = \sin\varphi_0$ is the initial, normalized, current amplitude. From this equation the value of $\eta = I/I_c$ for which the escape to the voltage state in a Josephson junction occurs can be easily evaluated as the value $\eta_E$ for which $\Delta U_E$ goes to zero. A comparison of such a calculation with the numerical data is shown in Fig. 8.

In Fig. 8a we have compared the analytical prediction obtained from (4) with the numerical data for values of $\dot\eta$, the current sweep rate in Josephson terms, varied over three orders of magnitude setting $\alpha = 0$ (the analytical calculation, in principle, is valid in this limit): we see that the agreement between equation (4) and numerical data is very good for $0 \leq \varphi_0 < 1$. In Fig. 8b instead we investigate, for a fixed value of the sweep rate, $\dot\eta = 1.8 \times 10^{-6}$, the effect of different values of the dissipation on the escape current. We see here that the agreement between the predictions and the numerical is very good when $\alpha$ is lower than the sweep rate, the condition in which we are observing the relevance of the initial conditions. This confirms the relevance of the parameter $\kappa$ and the physical nature of the phenomenon generating early escapes which are due to the extra energy pumped in the system by the initial condition when $\kappa < 1$. In these conditions we see that the proposed lowering of the potential predicted by eq. 4 gives a very good prediction of the escape current. Note in Fig. 8b that, for a given value of the sweep rate, when a "saturation" value of $\alpha$ is reached (below the normalized sweep rate) further lowering of it does not produce relevant changes in the escape current. For high values of the loss parameter, as epitomized in Fig. 5 there are no changes of the escape currents and the escape value always equals 1.

## 3) NON-ZERO TEMPERATURE

Let us step now to the analysis of the escape process for nonzero temperatures. Thermal fluctuations are plugged into the right hand side of eq. 2 by the term $n(t)$ linked to the dissipation through the fluctuation-dissipation relationship [18]: $<n(t)> = 0$ and $<n(t)n(t')> = 2\alpha \frac{k_B T}{E} \delta(t-t')$, where $T$ is the thermodynamic temperature of the system, $k_B$ =1.38 x10$^{-23}$ J/K is Boltzmann's constant and $E$ an adequate normalizing energy (in Josephson effect terms this is just the $E_J$ introduced in the past section). In Fig. 9a we show three statistical escape distribution obtained for five values of the parameter $\frac{k_B T}{E}$ which were, from left to right respectively $1.8x10^{-2}$, $1.3x10^{-2}$, $8.0x0^{-3}$, $4.0x10^{-3}$, $2.0x10^{-3}$, $5.0x10^{-4}$. These distributions were obtained by setting the initial conditions $\varphi_0 = 0$ and $\dot\varphi_0 = 0$, with standard deviation around these values given by $(k_B T/E)^{1/2}$, and setting $\alpha$ =0.018 and $\dot\eta = 1.8x10^{-9}$.

Technically, the distributions of Fig. 9a are obtained generating 1500 escape events by increasing $\eta$ and recording each time the value of it for which the escape occurs : on the vertical axis we just have the number of events (switches) relative to the specific $\eta$ value. The horizontal resolution is $6.4 x10^{-5} < \Delta\eta < 5.3x10^{-4}$ and it is varied depending on the width of the specific statistical distribution. On the vertical axis we have essentially a probability distribution $\rho$ expressed in arbitrary units. The distributions move right with temperature and the width of the statistical distributions decreases with temperature. In Fig. 9b we show the temperature dependence of the widths of the statistical distributions for the same value of $\dot\eta = 1.8x10^{-6}$ but for five different values of the loss factor $\alpha$ which are one order of magnitude apart starting from $1.8x10^{-4}$ down to $1.8x10^{-8}$. We see that the dependencies are straight lines in a log-log plot (here we consider only the data returning linear correlation coefficients above 0.999) and the values of the slope return the value of the exponent $\gamma$ of the power law indicated in the inset which varies in the interval *[0.59;0.70]*. Kramers theory (see eq. 12 of ref. 11, the paper is devoted to Josephson phenomenology) predicts a dependence $\sigma \approx (k_B T/E_J)^{2/3}$ of the width of the statistical distributions on temperature; we see that only one straight line of the log-log plot returns a value close to *2/3* while the other values are different up to *10%* from this value.

In Figure 9c we show the dependence of the slope of the straight lines of Fig. 9b, namely the exponent $\gamma$ of the law $\sigma = (\frac{k_B T}{E})^\gamma$, upon the parameter $\kappa$. We see that the slope of the curves has a maximum around $\kappa=1$ and is roughly symmetrical around this point. Thus, the dynamical response changes significantly as a function of the ratio a $\frac{\alpha}{\dot\eta}$ and, since the value of the sweep rate is fixed, the result indicates that is the damping parameter which regulates the response of the system. Analogous results can be obtained varying $\dot\eta$.

In Fig. 10a we show a plot similar to that of Fig. 9b obtained now setting a given angle as initial condition, but still initial angular velocity set to zero, namely $\varphi_0 = 0.2\pi$ and $\dot\varphi_0 = 0$. For the two lower values of the loss parameter (*1.8x10⁻⁷* and *1.8x10⁻⁶*) we obtain the same value of the exponent of the power law $\gamma$, within few parts over *10³* uncertainty, and this value is equal to *0.5*. The fitting for the highest loss value instead (*1.8x10⁻⁵*) returns the exponent *0.672*. In Fig. 9b we show the dependence of the exponent $\gamma$ of the law $\sigma = (\frac{k_B T}{E})^\gamma$ upon $\log\kappa$. We see here, more clearly than in Fig. 9c, that the response of the system exhibits abrupt differences below and above $\kappa = 1$. It is worth noting that the same dependence shown in Fig. 10b, is recorded if we set a different angle as initial condition: along with $\varphi_0 = 0.2\pi$ we tested the results for and $\varphi_0 = 0.5\pi$ and $\varphi_0 = 0.1$ and these cases returned exactly the same dependence in the range of $\log\kappa$ shown in Fig. 9b . We conclude that the effect of setting a given angle as initial condition is to generate, for $\kappa < 1$, a straightforward dependence of the standard deviation of the statistical distributions upon the square root of the temperature $\sigma = \sqrt{\frac{k_B T}{E}}$ valid at least over two orders of magnitude of the parameter $\frac{k_B T}{E}$.

If we now choose as initial condition an angle distributed randomly and uniformly in a given interval $[-\varphi_0;\varphi_0]$, the above found square root dependence of the standard deviation of the distributions upon $k_BT/E$ fails. What happens in this case quantitatively evident in Fig. 11a,b,c where we show the dependence of the standard deviation of the distributions upon the normalized thermal energy in a log-log plot. Here we see that for lower values of the normalized energy a saturation of the standard deviation occurs and the specific saturation current depends on the initial angle. The result are obtained setting an initial angle randomly uniformly distributed respectively in the intervals *[-0.1;0.1]* (a) , *[-0.2π;0.2π]* (b) and *[-π/2;π/2]* (c). In the plots we also show the dependence of the "saturation" curves upon the loss parameter. Looking back at the *T=0* results we can have a straightforward explanation of this effect considering, in particular, Fig. 6b. We see there that different initial angles correspond to different escape currents and therefore if the initial angle is uniformly distributed over an interval the escape $\eta_E$ also shall be distributed over an interval which will set a distribution even for *T=0*. When the temperature in the system is high enough these initial condition-generated oscillations shall not be visible because are masked by thermal fluctuations, however, when thermal fluctuations have a low energy the standard deviation is just determined by the randomness of the initial angle distribution.

In Figs. 11a,b,c we also write the slopes of the "linear" portions of the plots, those corresponding to higher temperatures. It is worth noting that results similar to those of Fig. 10, obtained setting an initial uniform random distribution of the angle can be obtained even with different type of random initial conditions.

The above argument concerning the saturation of the standard deviation of the escape distributions based on the *T=0* behavior, makes sense when $\kappa$ is of the order of the unity and below. When $\kappa$ is much above the unity the distribution of escape values does not depend on the initial angle, as we see in Figs. (3-7), and therefore saturation cannot be expected on the basis the above conjecture. However, in the nonzero temperature case variations of the plots of Figs. (3-7) can be expected and saturations effects are recorded even at very low temperatures even for *κ>1*, as shown in Fig. 11. We see in this figure that the "saturation" $\sigma$-value of statistical distribution at low energies/temperatures depends on the amplitude of the random initial angle interval and on the value of the loss parameter.

We have seen in Fig. 11 that, for a given value of $\dot{\eta}$ the with of the distributions can attain a saturation value that will call $\sigma_{SAT}$. In Fig. 12 we show the dependence of this parameter upon the loss parameter $\alpha$ in a semi-logarithmic plot for three settings of the initial conditions corresponding to have the angle uniformly distributed respectively in *[-0.1; 0.1]*, *[-0.2π;0.2π]* and *[-0.5 π ; 0.5 π]*: Fig. 12a has been obtained for $\dot{\eta} = 1.8 \ x \ 10^{-6}$. Instead, in Fig. 12b we show the same dependencies of the saturation currents, for the same initial conditions, obtained for $\dot{\eta} = 1.8 \ x \ 10^{-7}$. We clearly see that both 12a,b that the dependencies are exponential of the type $\sigma_{SAT} = \sigma_0 e^{-s\alpha}$ with slopes *s*, indicated in the insets, which have only few percent differences between the two cases (for the same random initial interval) . For both cases we see that the extrapolation of the straight lines toward *α = 0* is a well defined, non zero, value and therefore, for a random initial angle the width of the distribution has a finite value in a lossless system. "Parental" results of those shown in Fig. 12a,b were recently obtained in ref. 14 for the width of the peaks appearing in the statistical distributions for high sweep rate: in that case the dependence of the width of the individual peaks of a multipeaked distribution was investigated and it was shown that the dependence of the width of the peaks on the loss factor was exponential.

## 4) EXPERIMENTAL BACKGROUND

We note that the work herein contained is referred to very adiabatic changes of force terms in the system (2). The effects of "pulsed" excitations on this system have been thoroughly investigated in previous publications (see [10] and references therein). The pulsed experiments on our systems, and the oscillations these generated are much reminiscent of the phenomenology early observed in pulsed laser experiments [19]; in the present paper, however, we have undertaken a systematic analysis of the "adiabatically" driven system (2) since we believe this effect can serve as useful background of knowledge for investigating more complex and non adiabatic conditions. This was the case indeed of the work recently appeared in ref. 14 for which a guide from (preliminary) results herein presented was important.

Framing of our results in terms of experimental parameters and procedures can be illustrated for a specific case. Systems which have represented a benchmark for the one-dimensional potential herein treated are Josephson junction and interferometers (consisting of a superconductive loops closed by one or two junctions). The first thing to point out is that the operating temperatures at which Josephson systems have been typically operating have gone from *4.2K* down to *10mK*. What we have used as normalizing energy $E$ is, in Josephson terms, the zero-bias Josephson energy which is the $E_J$ defined earlier in sect.2. Now, typical maximum pair currents for the thermal activation experiments performed on Josephson junctions are in the range *(1-330)μA* [20-26]. If we consider for example a *1μA* Josephson current, the corresponding Josephson energy is $E_J = 0.33 \times 10^{-21}J$ and our normalized $k_BT/E_J$ will be *0.16* at *4K* and *4x10⁻⁴* at *10 mK*. When the Josephson maximum current scales up one order of magnitude the energy will scale up of the same amount and the maximum and minimum values of the normalized energy will consequently scale down a factor *10*. Thus, for a Josephson current of *100 μA* the energy $E_J = 33 \times 10^{-21} J$ and the normalized temperature will range from *0.16 x 10⁻²* at *4 K* down to *4x10⁻⁶* at *10 mK*.

From the above estimate of currents and related normalized energy it is evident why we have chosen the intervals of energy variation for our investigation. Other relevant parameters in our work are the rate of change of the applied force $\dot{\eta}$ and the loss factor $\alpha$; as far as the first parameter is concerned, also known in Josephson terms as normalized sweep rate since it is the normalized time derivative of the bias current, this can vary from the range of $10^{-6}$ down to the range of $10^{-9}$, depending on junctions properties (essentially capacitance and critical current) [20-25]. A special problem instead is represented by establishing the effective losses in Josephson systems when performing potential escape measurements; in this case the work reported in ref. 27, for example, provides interesting information because it evaluates a specific temperature dependence of the dissipation in the junctions and simulations should take into account this physical condition. An analysis of the intriguing effects generated by the variations of this parameter would be interesting and requires further investigations.

In the present work we have preferred to set constant the dissipation parameter $\alpha$ in order to maintain our analysis more general because in many physical cases the dissipation coefficient can just be kept constant. As we have seen at the beginning setting a constant dissipation factor and all zero initial conditions there are no effects on the escape from the potential and the data indeed might follow very accurately the Kramers model. In Fig. 13 we plot the numerically obtained position (in current) of the statistical distribution peaks with initial phase and phase derivative set to zero, for $\alpha=0.01$. The full circles represent the numerical results for the position in current of the statistical peak of the distributions, in values normalized to $I_c$ while the continuous line represents the prediction of Kramers theory [28]. According this theory the position, in current ($I_p$), of peak of the statistical distribution moves away from the $T=0$ Josephson critical curent $I_c$ following the same functional dependence on temperature of the width of the distribution, namely $(I_c - <I_p>)/I_c \approx (k_B T/E_j)^{2/3}$. A numerical evidence of the fact that peak widths and position follow the same temperature dependence is given in Fig. 51, right panel, of ref. 10 : as we see there, for the given value of the loss, the data follow perfectly Kramers predictions all over the temperature range.

However, if the loss factor $\alpha$ is scaled down to $10^{-10}$, namely below $\dot{\eta} = 2 \times 10^{-9}$, so that the parameter $\kappa = \frac{\alpha}{\dot{\eta}}$ is below the unity, at low temperatures the position of the peaks freezes as shown by the yellow triangles in Fig. 13. For this specific run we have chosen to set "Gaussian" initial conditions meaning that the initial angle is randomly distributed around the given angle $\varphi_0 = 0.0856\ \pi$ with a standard deviation of the packet proportional to the square root of the $k_BT$ ; the initial phase derivative is always set to zero. The triangles, the squares and diamonds in Fig. 13 correspond to the time steps of the numerical integration indicated in the inset, showing stability of the results, for very low dissipation, under "halving" tests. The intersection of horizontal line obtained for $\alpha = 10^{-10}$ with the Kramers curve occurs for a temperature slightly above *50 mK*. We can say that an initial angle $\varphi_0 = 0.086$, and a loss factor less than the normalized sweep rate, establish a difference between the temperature intervals in which the system responds according to the Kramers model or to the phenomenology of a very underdamped dynamical regime. As attempt frequency for tracing the theoretical Kramers curve, based on eq. 1, in Fig. 13 we have used both the one corresponding to the harmonic approximation, $f = f_{J0}(1-\eta^2)^{1/4}$ (where $f_{J0} = \frac{\omega_0}{2\pi}$ is the zero-bias Josephson plasma frequency [9,10]) and the one coming from the anharmonic approximation (see pag. 5 of ref. 10) but the two results are identical within *0.1%*.

We note that, according to the analysis performed in ref. 27, in a temperature interval between *1K* and *5K* the effective damping can change of *5* orders of magnitude and therefore it is not unreasonable to expect that in the hundreds of millikelvin range the normalized damping could be of the same order of magnitude, or less, of the sweep rate and generate the freezing of the position peaks. Since it has been shown that problems might exist [29,30] in attributing to quantum phenomenology and theories experimental results at very low temperatures, we speculate that more quantitative insight into the Josephson potential escape phenomenology could come from specific analysis of the resistively and capacitively junction model (RCSJ) of Josephson systems considering adequate settings for the effective losses and initial conditions. We note that in very underdamped conditions oscillating transients can hardly be removed, even imposing waiting times between current ramps.

## 5) CONCLUSIONS

We have investigated systematically the features of potential escape for a physical system having relevant impact in general physics and condensed matter. Although work has been dedicated in the past to identify analytical criteria to describe the response of the system that we have investigated and peculiar models have been considered to explain experiments [20-26] the nonlinearity of the potential, and the related dynamical equation it generates, are such that only accurate and systematic numerical integrations can enable us to have ideas of the response over wide parameters excursions at very low temperatures. The zero temperature analysis of the system (2) led us to identify a relevant parameter, the ratio between loss factor and rate of increase of applied force: a threshold related to it $(\kappa = \frac{\alpha}{\eta} = 1)$ characterizes the response of the dynamical system. This threshold differentiates the dynamical response both in zero temperature and in non-zero temperature. Below this threshold the dynamics is much dependent on the initial conditions imposed on the dynamical variable of the potential because the oscillations generated by the initial data condition the potential escape process. Moreover, we have shown that relevant deviations from the predictions of the Kramers model occur due to the setting of specific random initial data.

The present paper was also motivated by the fact that in a previous publication [30] evidence was found that macroscopic quantum tunneling theory cannot provide explanation of experimental results obtained on Josephson systems at very low temperature. We have herein shown that specific settings of the RCSJ model, as far as Josephson phenomenology is concerned [9,10], can return results which are close to the observed experimental phenomenology in terms of saturation of the statistical distributions and freezing of the position of the peak of the distributions, The physical motivation providing agreement between modelling and experiments is essentially the fact that in rather underdamped classical systems the effects of transients oscillations cannot be neglected.

The results herein presented are consistent with recently reported investigations of potential escape phenomena in which the effect of very high rates of change of the external force on the system (2), with noise term, was investigated [14]. In ref. 14 it was shown that the threshold $\kappa = 1$ represents a crucial condition for generating multipeaked statistical distributions. It is evident then that the behavior of the system for *T=0* is a background conditioning the response when thermal excitations are present. It will be surely interesting to further develop specific issues related to the Josephson effect and in particular those conceerning the temperature dependence of the dissipation parameter. Previous works indeed have attempted to find general criteria for the appearance of specific phenomena in the escape process [31].

It is also worth noting that presently much attention is devoted to digital circuits based on Josephson effect which could work without the shunting resistors necessary for the proper operation of Rapid Single Flux Quantum (RSFQ) logics [32, 33, 34]. In these conditions, in which the effect of loss is pushed to an extremely low limit, the response of the Josephson junctions to external excitations leading it to fast switches is a crucial phenomenon and the effect of the initial state of the system is even more relevant than it is for shunted junctions [35]. The same type of problems are faced when considering unshunted Josephson junctions as switches for single photon detectors [26, 36, 37], a field which has a promising impact both on fundamental and applied sciences. The relevance of our work can be particularly realized when considering the issues of ref. 26. It is important in this specific case to set the optimal parameters for which the response of the devices is not conditioned by internal modes or spurious oscillations.

**Acknowledgement**

This work was partially supported by the INFN-FEEL project (Italy).

# FIGURE CAPTIONS

**Figure 1**. The process of lowering the potential barrier of the system (2) by the external forcing term $\eta$. The inset shows the dependence of the amplitude of the potential (the difference in energy between the maximum and the minimum) as a function of $\eta$. The full diamonds and the empty circles indicate respectively the stable and unstable points of the potential. Escape from the well occurs when stable and unstable points coincide.

**Figure 2**. Traces of the minimum (continuous curve) of maximum (dashed curve) of the well in Fig.1 traced as a function of $\eta$ for "flat" initial conditions, namely $\varphi = 0$ and $\dot{\varphi} = 0$. As indicated in the plots, a) and b) correspond to loss factors $\alpha$ which are *6* orders of magnitude apart. Escape occurs when the two curves cross and we see that, in spite of the noticeable difference in loss, it occurs in both cases for *η=1*. In this integration.

**Figure 3**. The changes in stability and equilibrium generated lowering loss, setting as initial conditions on the phase $\varphi_0 = 0.3\pi$, $\dot{\varphi}_0 = 0$. We can see in a) that for $\alpha = 0.01$ the response is identical to that of the panels of Fig. 2. However, decreasing the loss $\alpha$ parameter to *3.2x10⁻⁷* in b), and to *3.2x10⁻⁸* in c), the $\eta$ point for which the two curves cross becomes less than the unity, since the initial oscillations due to the initial angle do not damp out. The fine structure of the dark areas is evident in the magnifying zoom showing the phase oscillations and the crossing with the instability curve.

**Figure 4**. Same as in Fig. 3 for initial conditions $\varphi_0 = 0.6\pi$ and $\dot{\varphi}_0 = 0$. We see that now in b) and c) that the escape occurs values of $\eta$ lower than those recorded in Fig. 3.

**Figure 5**. 3D plot showing the dependence of the "escape" values of $\eta$ upon variations of loss term and initial angle. The plot is obtained for a sweep rate $\dot{\eta} = 1.95 \times 10^{-8}$.

**Figure 6**. Projections of the 3D plot of Fig. 5 on the ($\eta_e$, $\varphi$) plane (a) and on the ($\eta_e$, $\log\alpha$) plane (b).

**Figure 7** (a) Dependence of the escape $\eta_e$ upon the loss factor $\log \alpha$ with the sweep rate $\dot{\eta}$ set as a parameter; (b) "Normalization" of the data in (a) obtained by scaling the loss factor by the sweep rate. The singularity in the response of the system, as far as escape processes are concerned, occurs when the ratio $\kappa = \frac{\alpha}{\dot{\eta}} = 1$.

**Figure 8** (a) Comparison between the prediction of eq. 4 and the numerical simulations for three values of the sweep rate; (b) the results of eq. (4) compared with numerical data inserting loss in the system. We see that the effect of the internal oscillations due to the initial angle become relevant for $\kappa<1$.

**Figure 9.** (a) Typical dependence of the position of the central peak of the statistical escape distribution and of the width of the distribution itself upon temperature : the peaks move toward increasing current and the width of the distributions tends to squeeze; (b) dependence of the width of the distribution statistics upon temperature setting initial angle and phase derivative equal to zero; (c) the dependence of the slopes extracted from the plot (b) upon the ratio $\kappa$.

**Figure 10.** (a)The dependence of the width of the statistical distributions upon the temperature for non-zero initial angles. The values of the initial angles are indicated in the plot. In (b) we see the dependence of the slope upon the parameter $\kappa$ when the initial angle is fixed to *0.2 π*. We can clearly see that passing through the interval $0 < \kappa < 1$ the response of the system strongly changes.

**Figure 11 .** The saturation of the widths of the distributions for different values of the initial angle distribution. In (a), (b) and (c) we set as initial condition and angle uniformly distributed respectively in *[-01;0.1]*, *[-2π ;2π]*, and *[-5π ;5π]*. In each plot the different symbols indicate different values of the loss parameter indicated in the inset. In the insets we also indicate the "asymptotic" high-temperature value of the exponent (*γ*) of the law *σ =($k_B T/H_j$)$^\gamma$*. The dotted

straight lines indicate, for comparison, the behavior observed for a fixed angle: in this case we observe no saturation all over the investigated temperature range.

**Figure 12.** The dependence of the value of the distribution width saturation value $\sigma_{SAT}$ upon the loss parameter for different initial conditions. In the semi-log plot we can clearly see an exponential dependence of the saturation width upon the dissipation and in the insets we indicate the slope of the linear dependence. In (a) we have set a rate of increase of the force term $\dot{\eta} = 1.8x10^{-6}$ while in (b) $\dot{\eta} = 1.8x10^{-7}$.

**Figure 13**. The dependence of the current corresponding to the peak of the statistical distributions upon temperature as described by the Kramers model (continuous line) and the results of numerical simulations (blue full circles) obtained setting all zero initial conditions, $\alpha=0.01$ and $\dot{\eta} = 2.085 \times 10^{-9}$. The yellow triangles, the squares and the diamonds show the results obtained, at low temperature, setting $\alpha = 10^{-10}$ the same value of $\dot{\eta}$ but initial conditions set by $\varphi_0 = 0.0856 \pi$ (with Gaussian noise around it) and $\dot{\varphi}(0) = 0$. The *dt* in the inset are the values of the time steps set for performing halving checks for the numerical integration.

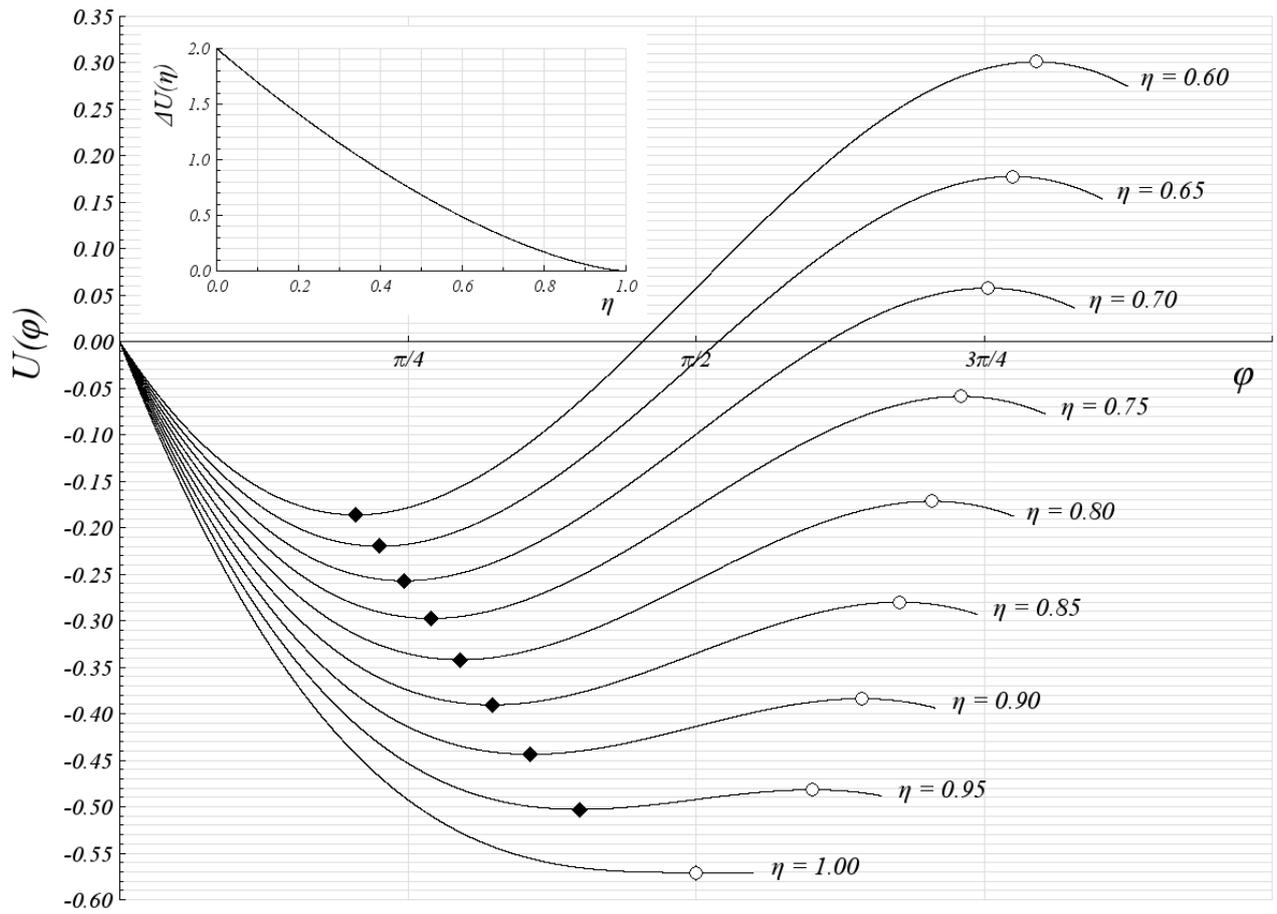

Figure 1

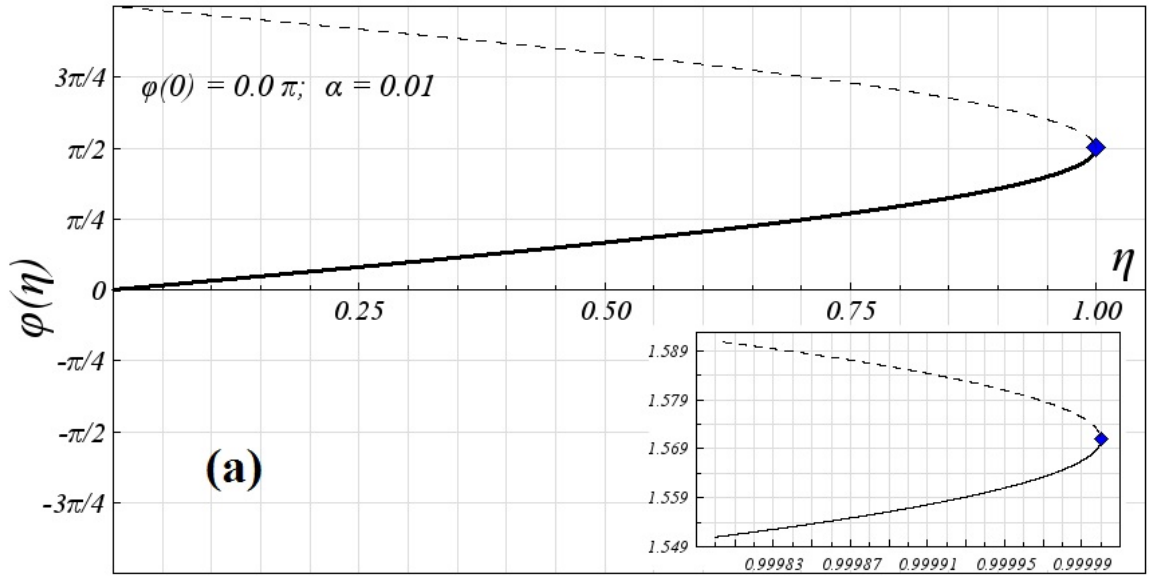

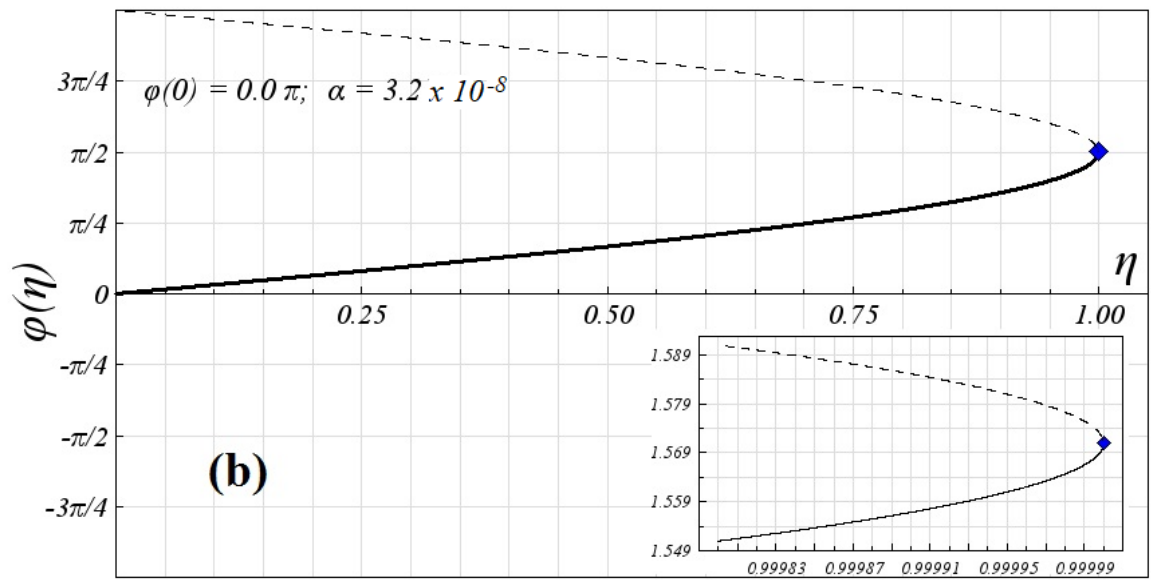

Figure 2

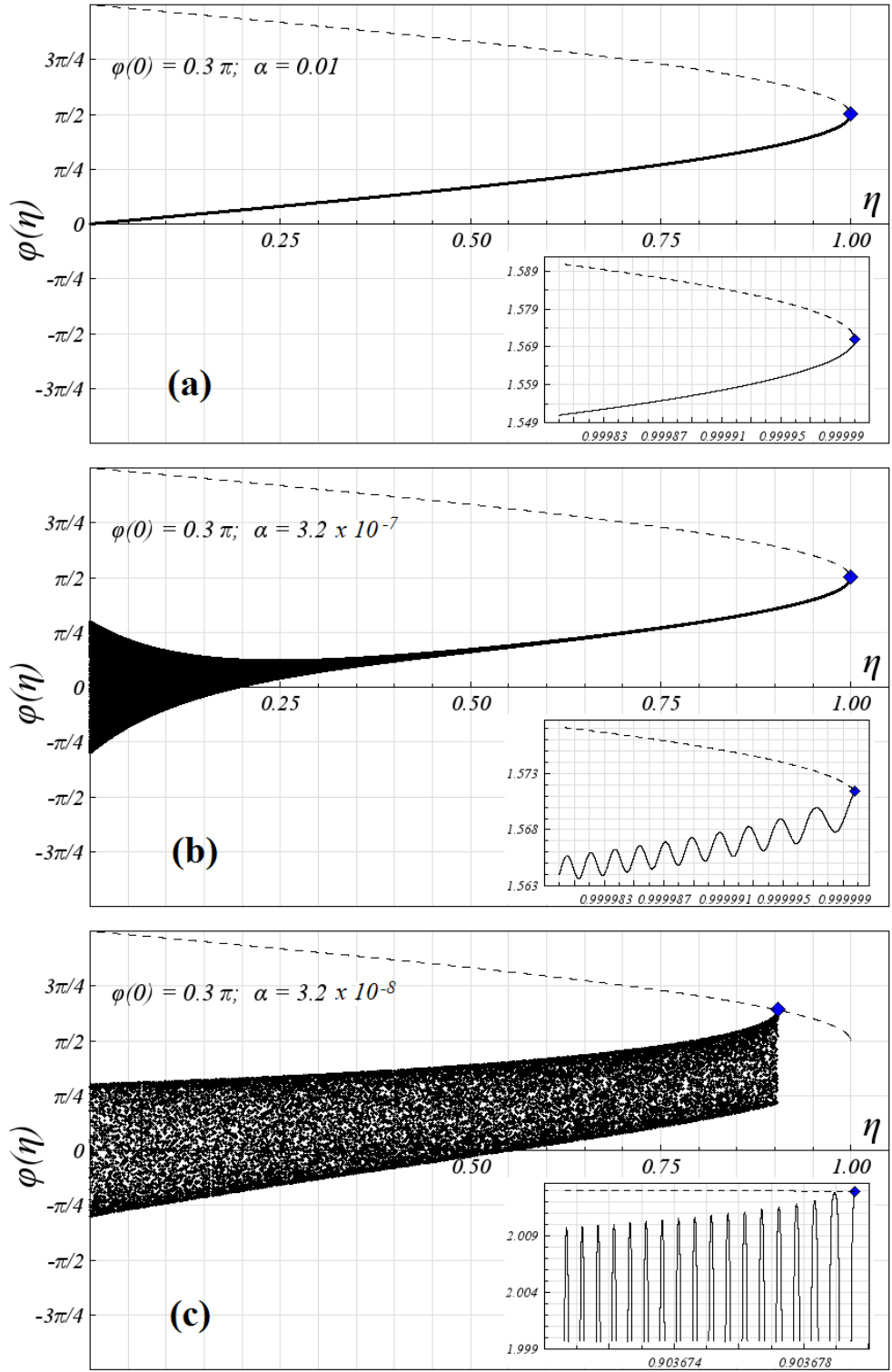

Figure 3

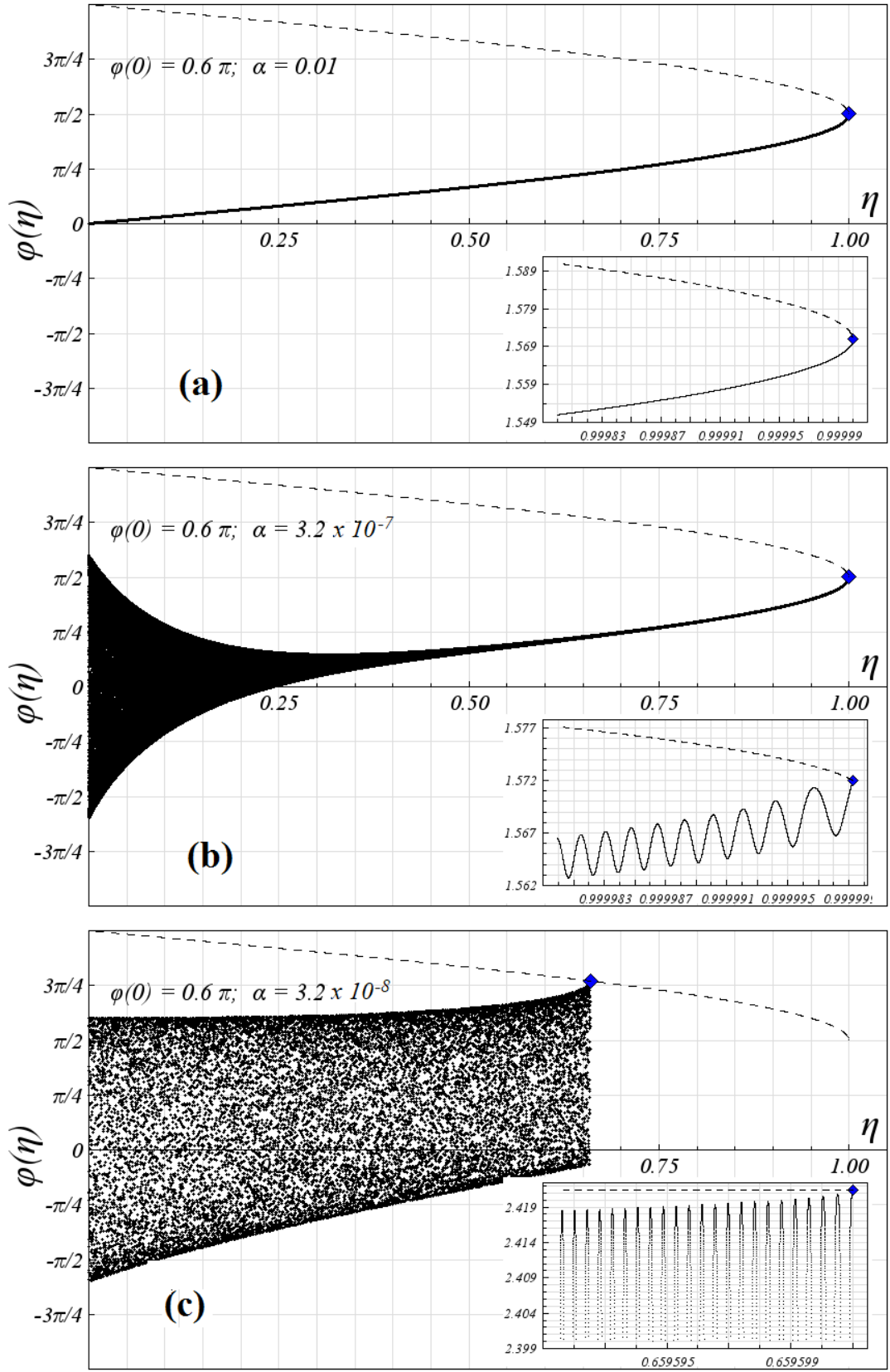

Figure 4

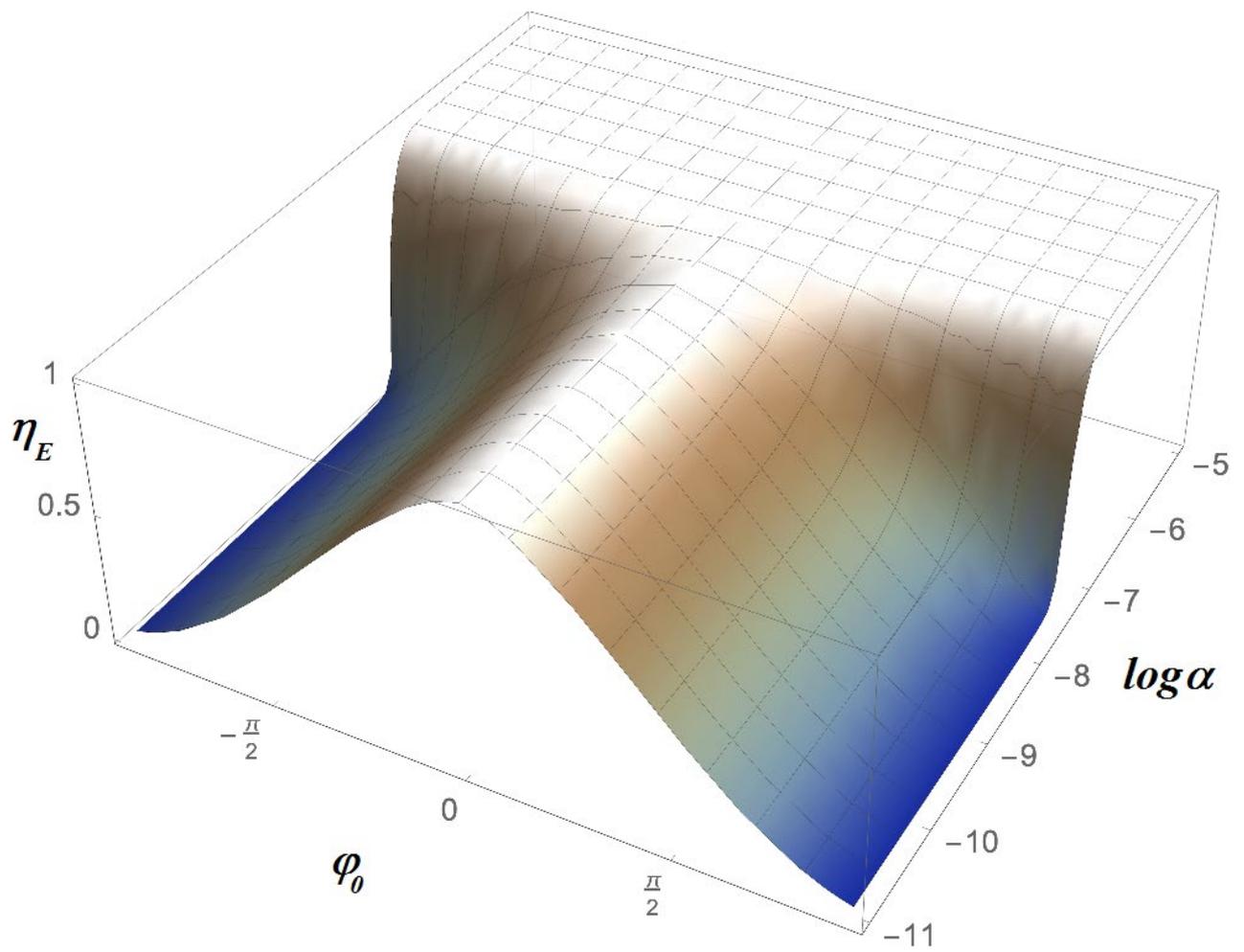

Figure 5

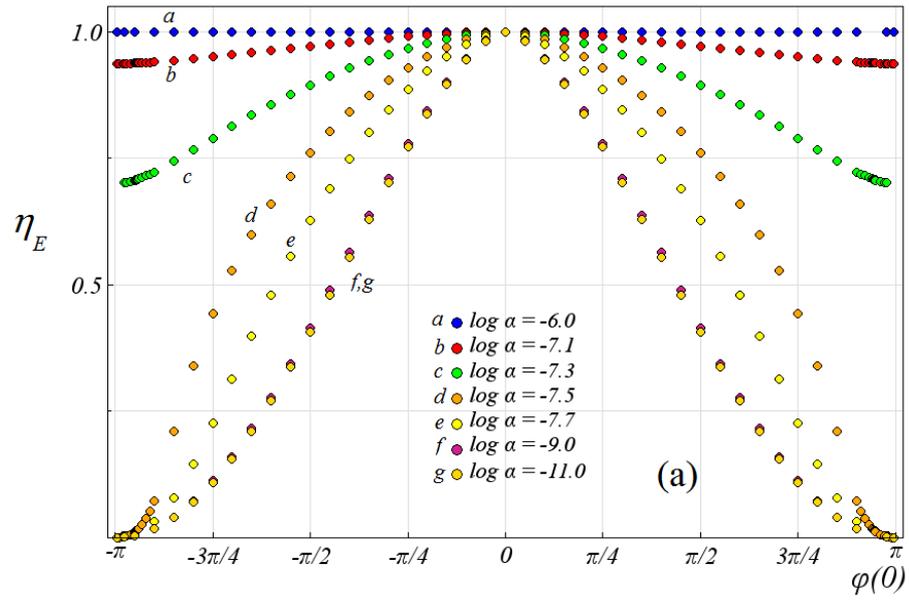

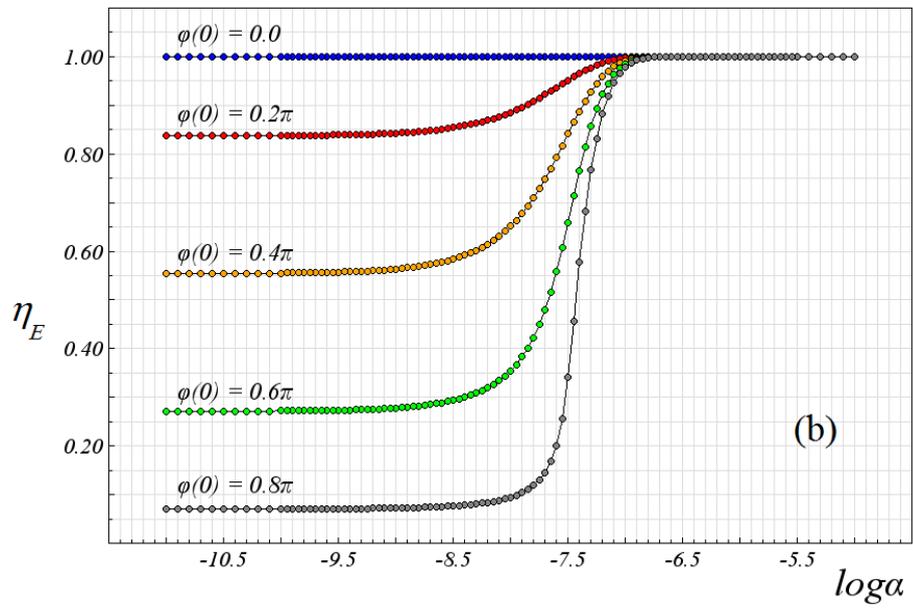

Figure 6

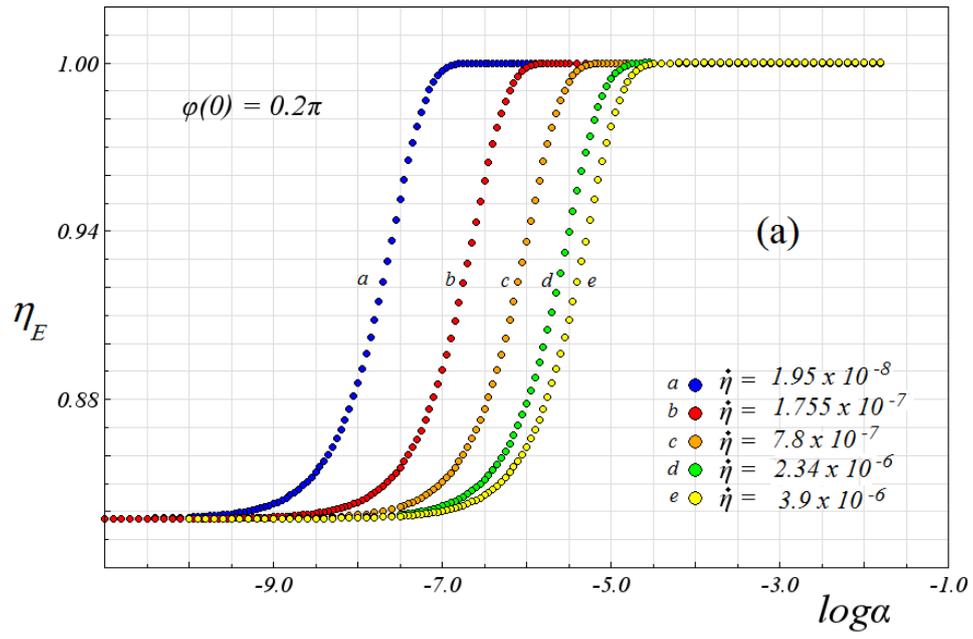

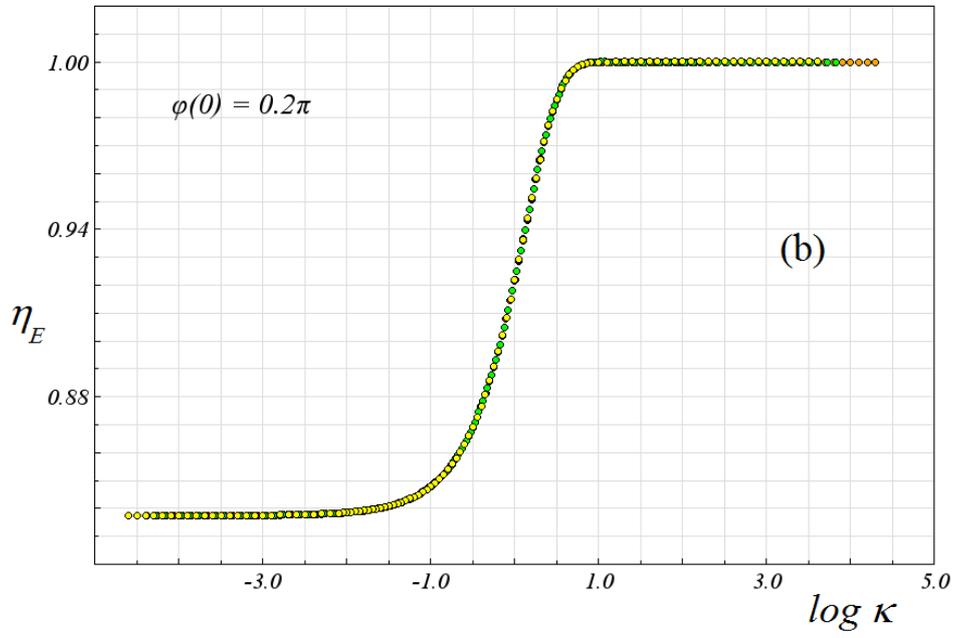

Figure 7

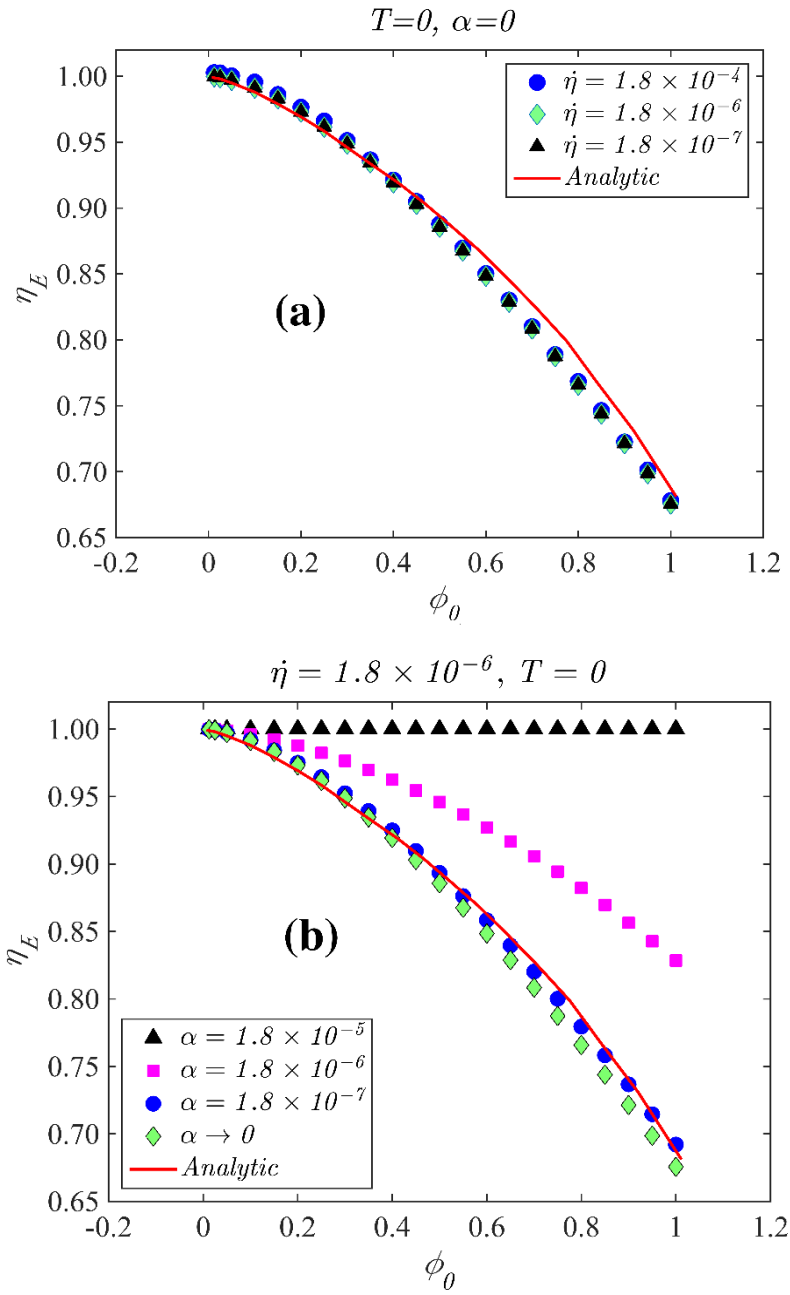

Figure 8

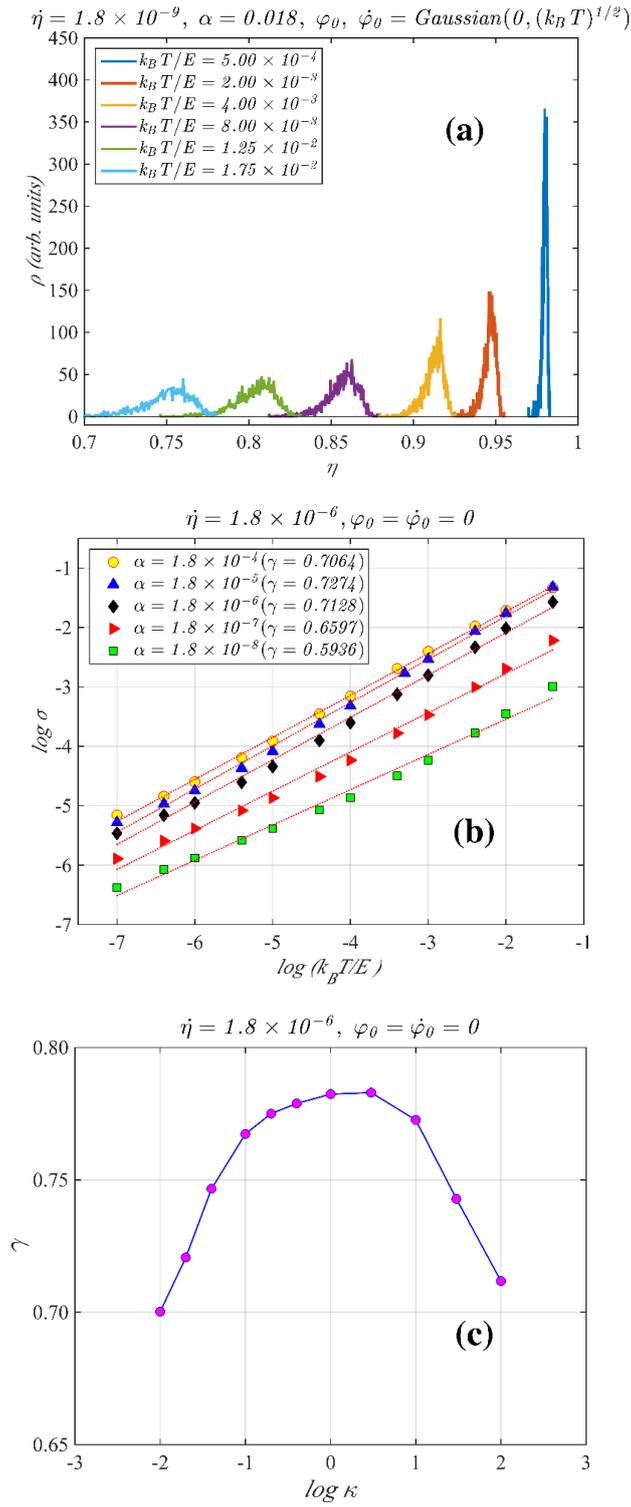

Figure 9

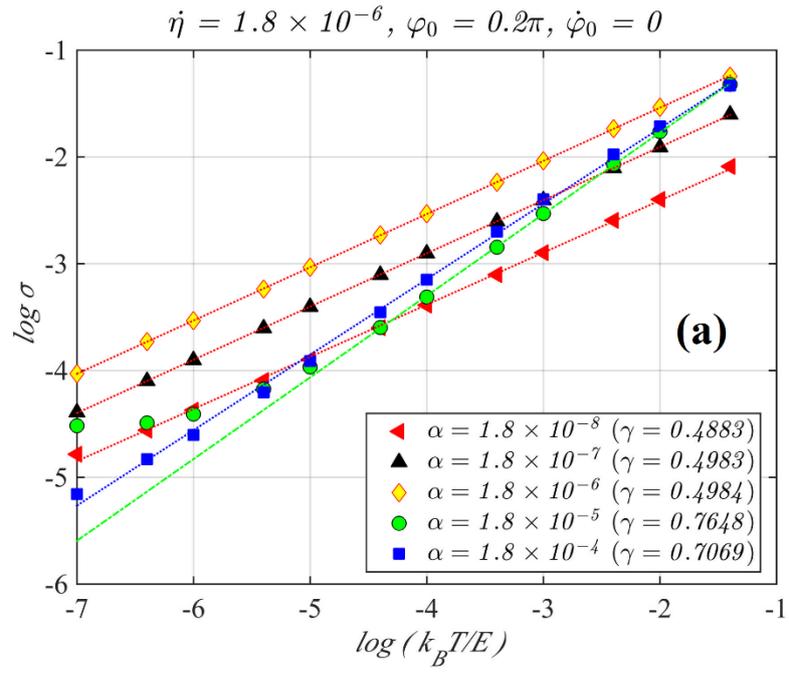
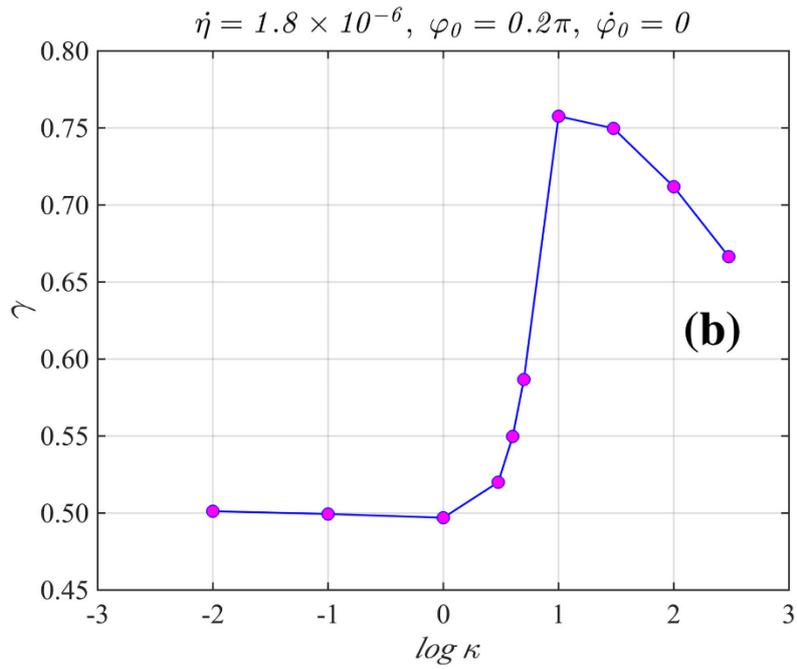

Figure 10

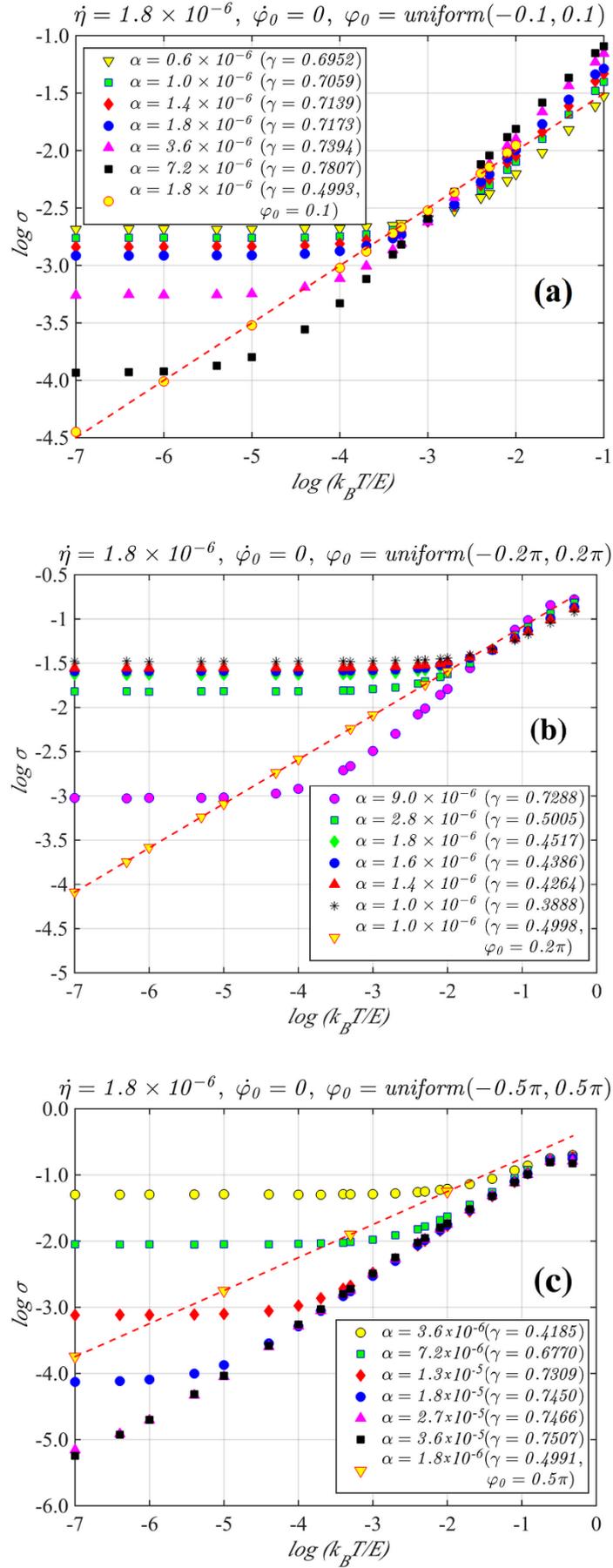

Figure 11

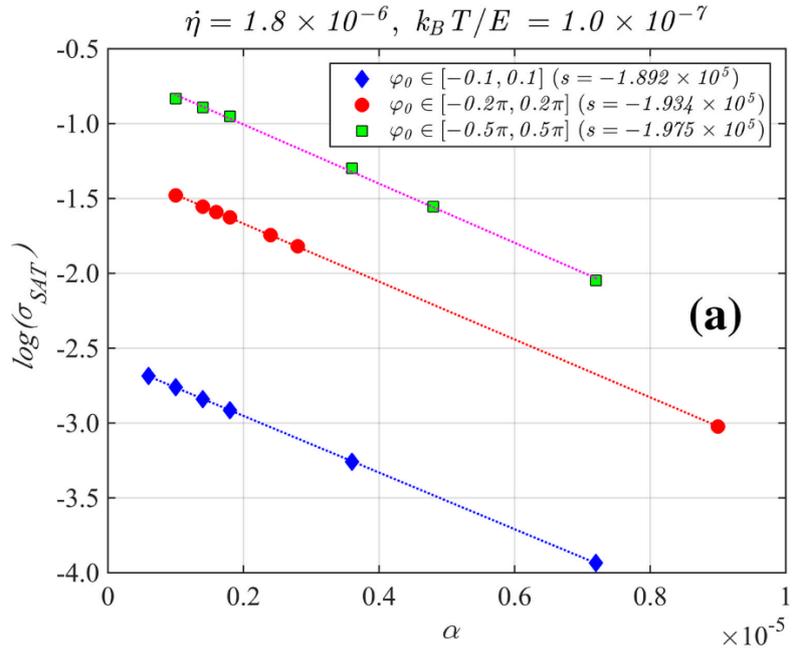

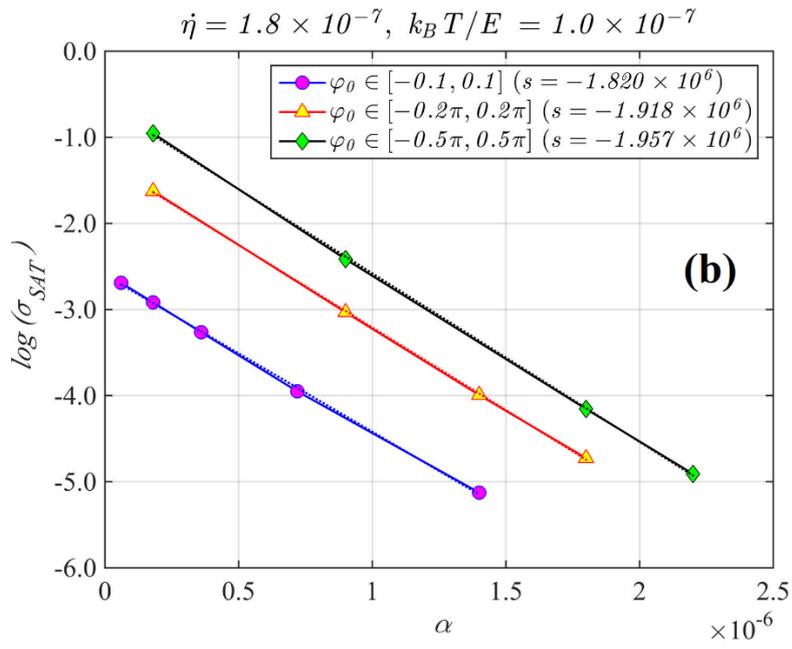

Figure 12

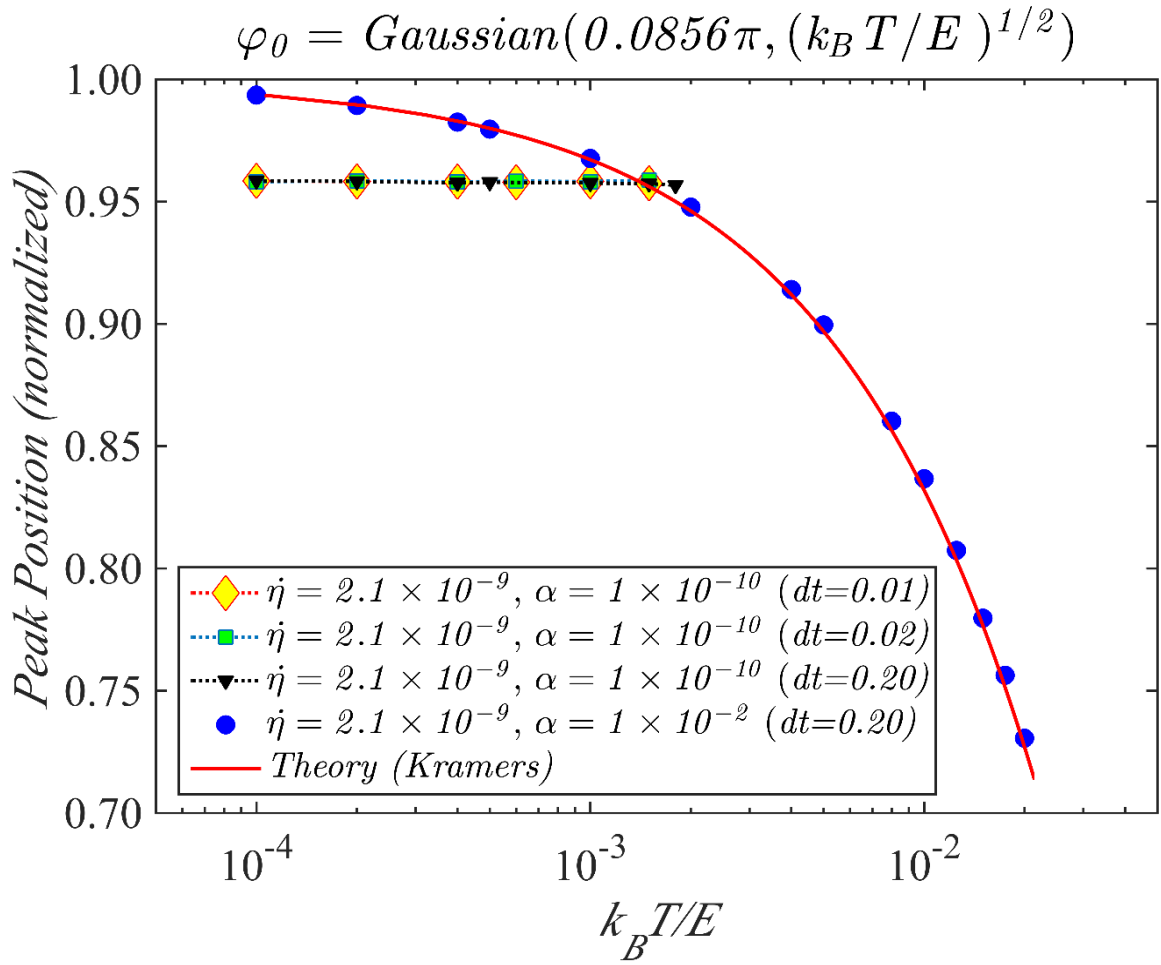

Figure 13